 \documentclass[11pt,a4paper]{article}
 \usepackage{amsmath}
\usepackage{cite}
\usepackage{graphicx}
\usepackage{amsfonts}
\usepackage{amssymb}
\input epsf

\def\br{\begin{eqnarray}}
\def\er{\end{eqnarray}}
\def\be{\begin{equation}}
\def\ee{\end{equation}}
\def\({\left(}
\def\){\right)}
\def\rlx{\relax\leavevmode}
\def\IR{\rlx\hbox{\rm I\kern-.18em R}}

\def\s{\sigma}

	\newcommand{\ba}{\begin{array}}
	\newcommand{\ea}{\end{array}}
	\newcommand{\beqa}{\begin{equation}\begin{array}{rcl}}
	\newcommand{\eeqa}[1]{\end{array}\label{#1}\end{equation}}

\def\IZ{\rlx\hbox{\sf Z\kern-.4em Z}}
\def\IR{\rlx\hbox{\rm I\kern-.18em R}}
\def\IC{\rlx\hbox{\,$\inbar\kern-.3em{\rm C}$}}
\def\one{\hbox{{1}\kern-.25em\hbox{l}}}

\renewcommand{\theequation}{\thesection.\arabic{equation}}
\def\IZ{\rlx\hbox{\sf Z\kern-.4em Z}}
\def\IR{\rlx\hbox{\rm I\kern-.18em R}}
\def\IC{\rlx\hbox{\,$\inbar\kern-.3em{\rm C}$}}
\def\one{\hbox{{1}\kern-.25em\hbox{l}}}
     
\begin{document}
     
\newcommand{\sect}[1]{\setcounter{equation}{0}\section{#1}}
\renewcommand{\theequation}{\thesection.\arabic{equation}}



     
     
     
     

\newcommand{\m}{ \mu}
\newcommand{\la}{ \lambda}
\newcommand{\C}{ C_0 }
\newcommand{\p}{ \varphi}

\newcommand{\ka}{ \kappa}

\newcommand{\e}{ \varepsilon}

\newcommand{\spa}{$\spadesuit$}
\newcommand{\bul}{$\bullet$}

\newcommand{\ir}{{\mathrm{IR}}}
\newcommand{\uv}{{\mathrm{UV}}}

\newcommand{\ml}{{\mathrm{ml}}}
\newcommand{\ms}{{\mathrm{ms}}}
\newcommand{\ns}{{\mathrm{n.s.}}}

\newcommand{\eff}{{\mathrm{eff}}}

\newcommand{\til}[1]{\tilde{#1}}

\def\PRL#1#2#3{{\sl Phys. Rev. Lett.} {\bf#1} (#2) #3}
\def\NPB#1#2#3{{\sl Nucl. Phys.} {\bf B#1} (#2) #3}
\def\NPBFS#1#2#3#4{{\sl Nucl. Phys.} {\bf B#2} [FS#1] (#3) #4}
\def\CMP#1#2#3{{\sl Commun. Math. Phys.} {\bf #1} (#2) #3}
\def\PRD#1#2#3{{\sl Phys. Rev.} {\bf D#1} (#2) #3}
\def\PRB#1#2#3{{\sl Phys. Rev.} {\bf B#1} (#2) #3}

\def\PLA#1#2#3{{\sl Phys. Lett.} {\bf #1A} (#2) #3}
\def\PLB#1#2#3{{\sl Phys. Lett.} {\bf #1B} (#2) #3}
\def\JMP#1#2#3{{\sl J. Math. Phys.} {\bf #1} (#2) #3}
\def\PTP#1#2#3{{\sl Prog. Theor. Phys.} {\bf #1} (#2) #3}
\def\SPTP#1#2#3{{\sl Suppl. Prog. Theor. Phys.} {\bf #1} (#2) #3}
\def\AoP#1#2#3{{\sl Ann. of Phys.} {\bf #1} (#2) #3}
\def\PNAS#1#2#3{{\sl Proc. Natl. Acad. Sci. USA} {\bf #1} (#2) #3}
\def\RMP#1#2#3{{\sl Rev. Mod. Phys.} {\bf #1} (#2) #3}
\def\PR#1#2#3{{\sl Phys. Reports} {\bf #1} (#2) #3}
\def\AoM#1#2#3{{\sl Ann. of Math.} {\bf #1} (#2) #3}
\def\UMN#1#2#3{{\sl Usp. Mat. Nauk} {\bf #1} (#2) #3}
\def\FAP#1#2#3{{\sl Funkt. Anal. Prilozheniya} {\bf #1} (#2) #3}
\def\FAaIA#1#2#3{{\sl Functional Analysis and Its Application} {\bf #1} (#2)
#3}
\def\BAMS#1#2#3{{\sl Bull. Am. Math. Soc.} {\bf #1} (#2) #3}
\def\TAMS#1#2#3{{\sl Trans. Am. Math. Soc.} {\bf #1} (#2) #3}
\def\InvM#1#2#3{{\sl Invent. Math.} {\bf #1} (#2) #3}
\def\LMP#1#2#3{{\sl Letters in Math. Phys.} {\bf #1} (#2) #3}
\def\IJMPA#1#2#3{{\sl Int. J. Mod. Phys.} {\bf A#1} (#2) #3}
\def\AdM#1#2#3{{\sl Advances in Math.} {\bf #1} (#2) #3}
\def\RMaP#1#2#3{{\sl Reports on Math. Phys.} {\bf #1} (#2) #3}
\def\IJM#1#2#3{{\sl Ill. J. Math.} {\bf #1} (#2) #3}
\def\APP#1#2#3{{\sl Acta Phys. Polon.} {\bf #1} (#2) #3}
\def\TMP#1#2#3{{\sl Theor. Mat. Phys.} {\bf #1} (#2) #3}
\def\JPA#1#2#3{{\sl J. Physics} {\bf A#1} (#2) #3}
\def\JSM#1#2#3{{\sl J. Soviet Math.} {\bf #1} (#2) #3}
\def\MPLA#1#2#3{{\sl Mod. Phys. Lett.} {\bf A#1} (#2) #3}
\def\JETP#1#2#3{{\sl Sov. Phys. JETP} {\bf #1} (#2) #3}
\def\JETPL#1#2#3{{\sl  Sov. Phys. JETP Lett.} {\bf #1} (#2) #3}
\def\PHSA#1#2#3{{\sl Physica} {\bf A#1} (#2) #3}
\def\PHSD#1#2#3{{\sl Physica} {\bf D#1} (#2) #3}

\begin{titlepage}
\vspace*{-2 cm}
\noindent
\begin{flushright}
\end{flushright}

\vskip 1 cm
\begin{center}
{\Large\bf New Massive Gravity Holography} \vglue 1  true cm

{G.M.Sotkov}$^{*}$\footnote {e-mail: sotkov@cce.ufes.br, gsotkov@yahoo.com.br}, {C.P. Constantinidis}$^{*}$\footnote {e-mail: cpconstantinidis@gmail.com} and  {U. Camara dS}$^{*}$\footnote {e-mail: ulyssescamara@gmail.com}\\

\vspace{1 cm}

${}^*\;${\footnotesize Departamento de F\'\i sica - CCE\\
Universidade Federal de Espirito Santo\\
29075-900, Vitoria - ES, Brazil}\\

\vspace{5 cm}

\end{center}

\normalsize
\vskip 0.5cm

\begin{center}
{ {\bf ABSTRACT}}\\
\end{center}

\vspace{0.5cm}

We investigate the holographic renormalization group  flows and  the classical phase transitions that occur in  two dimensional QFT  model dual to the New Massive 3D Gravity coupled to scalar matter. Specific matter self-interactions generated by quadratic superpotential are considered. The off-critical $AdS_3/CFT_2$ correspondence determines the exact form of the $ QFT_2$ 's $\beta$ -function  and the singular part of the reduced free energy. The corresponding  scaling laws and critical exponents characterizing  the RG fixed points as well as the values of the mass gaps in the massive phases are obtained. 

\vspace{0.5 cm} 

\end{titlepage}

\tableofcontents 
\setcounter{equation}{0}
\section{Introduction}
\label{intro}
The $AdS_{d+1}/CFT_d$ correspondence \cite{malda} provides holographic description  of  the $d=4$ $SU(N)$ supersymmetric large N gauge theories  and its off-critical $(a)AdS_5/QCD_4$ version is expected to solve the problem of the strong coupling regime of $QCD_4$ \cite{witt}. In this context the  \emph{two dimensional} case represents rather  ``non-physical" problem, which however is known to be of \emph{conceptual} importance. Since two dimensional (super) conformal group is infinite, the specific features of its unitary representations \cite{bpz} allow to exactly calculate all the anomalous dimensions  and the $n$-points correlation functions of all the primary and composite fields. Another  important fact of purely $2D$ nature is the existence of a vast variety of \emph{integrable} perturbations of the corresponding $CFT_2$'s \cite{zam}, as for example (super)sine-Gordon and the abelian affine (super) Toda models \cite{muss},\cite{sz}, whose $S$-matrices, mass spectra, form-factors and  some correlation functions are known exactly \cite{azz}. Apart from the practical use of all these $2D$ models in the description of real condensed matter systems \cite{bose}, the huge amount of available exact results also permits to realize non-trivial  self-consistency checks of the (eventual) validity of the off-critical $AdS_3/CFT_2$ correspondence even \emph{out} of its original superstring/supergravity/SUSY gauge theories frameworks.

 In what concerns the lessons one can learn about the corresponding realistic higher dimensional $d=4$ models, we should mention however one serious disadvantage when 3D Einstein gravity of negative cosmological constant is used as 3D ``bulk" gravity. Since it has no local degrees of freedom  its properties as well as the ones of its 2D dual are rather different from  the properties of corresponding $d+1=5$ versions. It is therefore interesting to study examples of the \emph{off-critical} $AdS_3/CFT_2$ correspondence based on appropriate \emph{extensions} of the Einstein 3D gravity, that have features  similar to the ones of 4D and 5D Einstein gravity such as ``propagating gravitons", non-trivial vacua solutions, etc. The simplest model of such extended 3D gravity is given by the following ``higher derivatives" action, called New Massive Gravity (NMG)\cite{1}:             
\begin{eqnarray}
S_{NMG}(g_{\mu\nu},\sigma;\kappa,\Lambda)&=&\frac{1}{\kappa^2}\int d^3x\sqrt{-g}\Big\{\epsilon R+ \frac{1}{m^2} {\cal K}-\kappa^2\Big(\frac{1}{2}|\vec{\nabla}\sigma|^2+V(\sigma)\Big)\Big\}\ \label{acao}\\
\cal{K}&=&R_{\mu\nu}R^{\mu\nu}-\frac{3}{8}R^2, \ \kappa^2=16\pi G,  \ \epsilon=\pm1\nonumber
\end{eqnarray}
Its linearisation describes massive graviton (of two polarisations) interacting with a scalar matter. One can consider the new $\cal{K}$ terms above as one loop counter-terms appearing in the perturbative quantization of 3D Einstein gravity. As it was recently shown by Bergshoeff, Hohm and Townsend (BHT) \cite{1} the above model \emph{in the absence of matter}, $i.e.$ $\sigma=const$, unlike the case of higher dimensional $D=4$ and $D=5$ Einstein gravities with one loop counter-terms added, turns out to be \emph{unitary} consistent (ghost free) and \emph{super-renormalizable} for the both choices $\epsilon=\pm1$ of the ``right'' and ``wrong'' signs of the $R$-term, under certain restrictions on the values of the cosmological constant $\Lambda=-\frac{\kappa^2}{2}V(\sigma^*)$ and of the new mass scale $m^2$.
 
The problem we are interested in  concerns the \emph{classical} critical phenomena that take place in the (euclidean) $QFT_2$'s  dual to NMG model (\ref{acao}) and more precisely  the phase transitions that occur in  2D \emph{ classical statistical mechanics} models  in infinite volume, whose thermodynamical limits  represent  models dual to NMG. According to the \emph{off-critical} $AdS_3/CFT_2$ correspondence \cite{gub} the domain wall solutions (DW's) of  3D  gravity models of negative cosmological constant  provide an alternative ``dual'' description of the renormalization group (RG) flows in specific 2D \emph{deformed} conformal field theories $CFT_2$'s, which can be described as as appropriate $CFT_{2}$'s (called $pCFT_2$'s) perturbed by marginal or/and relevant operators \cite{witt} :
\begin{eqnarray}
S_{pCFT_2}^{ren}(\sigma)=S_{CFT_2}^{UV}+\sigma(L_*)\int d^2x\Phi_{\sigma}(x^i)\label{eq28}
\end{eqnarray}
In the holographic RG approach\cite{VVB,4} the ``running'' coupling constant $\sigma(L_*)$ of $pCFT_2$ is identified   with the scalar field  $\sigma(z)$ and the RG scale $L_*$ -- with the scale factor $e^{\varphi(z)}$ :
\begin{eqnarray}
ds^2=dz^2+e^{\varphi(z)}(dx^2-dt^2),\quad\quad
\sigma(x^i,z)\equiv\sigma(z),\quad \ L_*=l_{pl}e^{-\varphi/2}\label{intro1}
\end{eqnarray}
of the DW's solutions of the NMG-matter model (\ref{acao}). Once  the pair of dual theories is established, the set of ``holographic rules'' \cite{witt,VVB,4} allows to deduce many of the important features of the \emph{quantum} $pCFT_2$ - as anomalous dimensions, fields expectation values, etc. - from the \emph{classical} DW's solutions of the corresponding ``bulk'' gravity \cite{1,2,3,nmg,8}. It should be mentioned however that the explicit construction of DW's solutions, even in the simplest case of Einstein gravity, is a rather difficult problem. It requires  the knowledge of an auxiliary function $W(\sigma)$ called superpotential that allows to reduce the corresponding  DW's gravity-matter equations to specific BPS-like $I^{st}$ order system. The generalization of the superpotential method\cite{8} to the case of NMG model (\ref{acao}) was recently introduced  in refs.\cite{3,nmg} :
\begin{eqnarray}
&&\kappa^2V(\sigma)=2(W')^2\Big(1-\frac{\kappa^2W^2}{2\epsilon m^2}\Big)^2-2\epsilon\kappa^2 W^2\Big(1-\frac{\kappa^2 W^2}{4\epsilon m^2}\Big),\nonumber\\
&&\dot{\varphi}=-2\epsilon\kappa W, \ \ \dot{\sigma}=\frac{2}{\kappa}W'\Big(1-\frac{\kappa^2W^2}{2\epsilon m^2}\Big)\label{sis}
\end{eqnarray}
where $W'(\sigma)=\frac{dW}{d\sigma}$, $\dot{\sigma}=\frac{d\sigma}{dz}$ etc.

  The present paper is devoted to the complete description of the holographic RG flows and of the classical phase transitions that occur in the  $pCFT_2$ dual to the NMG model with quadratic matter superpotential $W(\sigma)=B\sigma^2+D $. The critical exponents characterizing all the RG fixed points as well as the values of the mass gap in the massive phases are calculated.
    
\setcounter{equation}{0}
\section{CFT's data of NMG model}
\label{cftdata}

One of the main statements of the Holographic renormalization group \cite{VVB,4} is that the above scale-radial identifications (\ref{intro1}) determine the form of the RG equations:
\begin{eqnarray}
\frac{d\sigma}{dl}=-\beta(\sigma)=\frac{2\epsilon}{\kappa^2}\frac{W'(\sigma)}{W(\sigma)}\bigg(1-\frac{W^2(\sigma)\kappa^2}{2\epsilon m^2}\bigg),\quad\quad l=\ln L_*\label{rg}
\end{eqnarray}
as well as the explicit $\beta$-function  of  the dual $pCFT_2$ model  in terms of the superpotential $W(\sigma)$ only. Let us remember that all the (critical) $CFT_2$ data is given by the asymptotic behaviour  of the NMG's domain wall solutions \cite{nmg}. The two types of real zeros of this $\beta-$function : (a) $ W'(\sigma_{a}^*)=0$ and (b) $W^2(\sigma_{b}^*)=\frac{2\epsilon m^2}{\kappa^2}$ indeed coincide with (part of) the extrema i.e. $V^{'}(\sigma_{A}^*)=0$ for $A=a,b$  of the matter potential $V(\sigma)$. Hence new  purely NMG  i.e. type (b) critical points exist only in the case when $\epsilon m^2 >0$. By construction both -(a) and (b) critical points- describe  $AdS_3$ vacua $(\sigma_A^*,\Lambda^A_{eff})$ of the NMG model
\begin{eqnarray*}
ds^2=dz^2+e^{-2\epsilon\sqrt{|\Lambda_{eff}^A|}z}(dx^2-dt^2), \quad\quad A=a,b
\end{eqnarray*}
where the effective cosmological constants $\Lambda^A_{eff}$ are defined by the vacuum values of the  corresponding scalar
3D curvature:
\begin{eqnarray}
R=-2\ddot{\varphi}-\frac{3}{2}\dot{\varphi}^2\equiv8\epsilon(W')^2\left(1-\frac{\kappa^2W^2}{2\epsilon m^2}\right)-6\kappa^2W^2\label{eq8}
\end{eqnarray}
 $i.e.$  we have $R_{vac}=-6\kappa^2W^2(\sigma_{A}^*)=6\Lambda_{eff}^A$.
  These critical points are known to correspond to $II^{nd}$ order phase transitions occurring in $pCFT_2$ 
where it becomes conformal invariant. Therefore the critical behaviour of this 2D model is described by a  set of  $CFT_{2}$'s of central charges :
\begin{eqnarray}
 c_A =\frac{3\epsilon L_A}{2l_{pl}}\left(1+\frac{L_{gr}^2}{L_A^2}\right),\quad\quad L_{gr}^2=\frac{1}{2\epsilon m^2}\gg l_{pl}^2,\quad\quad \kappa ^2 W^2(\sigma^*_{A})=\frac{1}{L^2_{A}}\label{cc}
\end{eqnarray}
calculated in the  approximation of small cosmological constants, i.e. $l_{pl}\ll L_{gr}< L_A$ by the Brown-Henneaux  asymptotic method \cite{9} appropriately adapted to the case of NMG coupled to scalar matter  \cite{8}, \cite{nmg}.

It is natural to consider the \emph{quantum} (euclidean) $pCFT_2$ in discussion as describing the universality class of the thermodynamical (TD) limit of certain  2D \emph{classical} statistical models. We are interested in studying  the \emph{infinite volume} critical properties  of these statistical  models by using the Wilson RG methods. As it well known (see for example \cite{cardy}, \cite{muss}) they are characterized by the scaling laws  and the critical exponents  of their TD potentials as for example the ones $y_A = \frac{1}{\nu _A}$ related to the singular part (s.p.) of  the reduced free energy (per 2D volume) $F^A_s$, to  correlation length $\xi_{A}$  and to  $\Phi_{\sigma}(x_i)$'s correlation functions:
\begin{eqnarray}
&&F^A_s(\sigma)\approx \left(\sigma- \sigma_A^*\right)^{\frac{2}{y_{A}}}, \quad\xi_A\approx(\sigma - \sigma_A^*)^{-\frac{1}{y_A}} ,\nonumber\\
&&G_{\Phi}^A(x_{12},\sigma)=<\Phi_{\sigma}(x_1)\Phi_{\sigma}(x_2)>_A\approx \frac{e^{-\frac{|x_{12}|}{\xi_A}}}{|x_{12}|^{2(2-y_A)}},\label{sl}
\end{eqnarray} 
at the neighbourhood of each critical point $\sigma_A^*$. Once the $\beta-$function (\ref{rg}) is given, it completely  determines the scaling properties of TD potentials, correlation functions, etc. under infinitesimal  RG transformations as follows \cite{cardy}:
\begin{eqnarray}
 &&\beta(\sigma)\frac{dF_s(\sigma)}{d\sigma} + 2F_s(\sigma)=0,\quad\quad \beta(\sigma)\frac{d\xi(\sigma)}{d\sigma} =\xi(\sigma),\nonumber\\
 &&|x_{12}|\frac{dG_{\Phi}(x_{12},\sigma)}{d|x_{12}|}+\beta(\sigma)\frac{dG_{\Phi}(x_{12},\sigma)}{d\sigma} + 2(2+\frac{d\beta(\sigma)}{d\sigma})G_{\Phi}(x_{12},\sigma)=0\label{fs}
\end{eqnarray}
One can easily verify for example that the above critical exponents (related to the $\Phi_{\sigma}$ field scaling dimensions $\Delta_{\Phi}^{A}$) are given  by the  values of the $\beta-$functions derivatives:
\begin{eqnarray}
y(\sigma^*_{A})=2-\Delta_{\Phi}(\sigma^*_{A})=-\frac{d\beta(\sigma)}{d\sigma}\Big\vert_{\sigma=\sigma^*_{A}}\label{sd}
\end{eqnarray}
In our case (\ref{rg}) they have the following explicit form (for $W\neq0$)\footnote{the \emph{singular} points $\sigma_{s}$ such that  $W(\sigma_{s})=0$ (where $\beta-$function diverges) divide the coupling space in few independent regions.}:
\begin{eqnarray}
y_{a}=y(\sigma_{a}^{*})=\frac{2\epsilon W''_{a}}{\kappa^2W_{a}}\Big(1-\frac{\kappa^2W_{a}^2}{2\epsilon m^2}\Big),\quad 
y_b=y(\sigma_{b}^{*})=-\frac{4\epsilon(W'_{b})^2}{\kappa^2W_{b}^2}, \quad W_{b}^2=\frac{2\epsilon m^2}{\kappa^2}.\label{sdim}
\end{eqnarray}
Their 3D-geometry counterparts appear in the asymptotics   of  the matter  field $\sigma(z)$ of corresponding DW's solutions of NMG model (see ref.\cite{nmg}) :
\begin{eqnarray}
\sigma(z)\stackrel{z\rightarrow\pm\infty}{\approx}\sigma_{A}^* - \sigma_{A}^{0}e^{\mp2\Delta_A\sqrt{|\Lambda^A_{eff}|}z},\ 
\Delta_A=1+\sqrt{1-\frac{m_{\sigma}^2(A)}{\Lambda_{eff}^A}},\ m_{\sigma}^2=V''(\sigma_A^*),\label{asymp}
\end{eqnarray}
thus confirming the basic rule of $AdS/CFT$ correspondence\cite{witt} : the \emph{scaling dimensions} of 2D fields are determined by the 3D effective \emph{cosmological constants} $\Lambda_{eff}^{A}$ and by  the asymptotic $\sigma-$vacuum states\footnote{in the case of self-interactions the effective masses are defined around each of the extrema $\sigma_{A}^{*}$ of $V(\sigma)$, i.e. $\sigma_{A}^{* \ \pm}=\sigma(z\rightarrow\pm\infty)$ and therefore we have $m_{\sigma}^2(\sigma_{A}^{*})=V''(\sigma_{A}^{*})$}\emph{ masses} $m_{\sigma}^2(\sigma_{A}^{*})$ as follows:
\begin{eqnarray}
m_{\sigma}^{2}(\sigma_{A}^{*})=-\Lambda_{eff}^{A}y_{A}(y_{A}-2)\label{mas}
\end{eqnarray}
Depending on the values of $y_{A}$ (or equivalently of $m_{\sigma}^2(A)$) we can have three qualitatively different near-critical behaviours of the coupling constant $\sigma(l)$ and therefore different type of critical points determined by the dimensions of 2D fields $\Phi_{\sigma}$. As is well known when $\Delta_{\Phi} <2$ the corresponding \emph{relevant} operator gives rise to an  increasing RG flow away the (unstable) UV critical point, while for $\Delta_{\Phi} >2$ the operator governing the flow is \emph{irrelevant} and we observe decreasing RG flow towards the  (stable) IR fixed point :
\begin{eqnarray}
&&(UV)\quad\quad 0<y_{A}<2 ,\quad  m_{\sigma}^2(A)<0 ,\quad \quad  L_*\rightarrow 0, \quad\quad \xi\rightarrow \infty,\quad\quad   e^{\varphi}\rightarrow\infty,\nonumber\\    
&&(IR)\quad\quad y_{A}<0 ,\quad  m_{\sigma}^2(A)> 0, \quad\quad   L_*\rightarrow \infty, \quad\quad \xi\rightarrow 0,\quad\quad   e^{\varphi}\rightarrow 0 \label{UV}
\end{eqnarray}
 The ``degenerate" case  $y_A=0$, i.e. of (asymptotically) massless matter $m^2_{\sigma}(A)=0$, is known to describe marginal operators with $\Delta_{\Phi}=2$. Such critical points correspond to \emph{infinite order} phase transitions, characterized by an essential singularity 
$F^A_s(\sigma)\approx exp\left(\frac{\mu_{A}}{\sigma- \sigma_A^*}\right)$ and 
$\xi_A\approx exp\left(\frac{\rho_{A}}{\sigma- \sigma_A^*}\right)$ instead of the power-like scaling laws (\ref{sl}) for thermodynamic  potentials in the case of $II^{nd}$ order phase transitions. Negative $m^2_{\sigma}(A)$ (tachyons) for scalar fields in $AdS_3$ backgrounds  do not cause problems when  the Breitenlohner-Freedman (BF) condition \cite{BF}  :
\begin{eqnarray}
\Lambda_{eff}^A\le m_{\sigma}^2(A)\label{A3}
\end{eqnarray}
is satisfied. The unitarity of the purely gravitational sector of  NMG model (\ref{acao}) requires that the following   Bergshoeff-Hohm-Townsend (BHT) conditions \cite{1}:
\begin{eqnarray}
m^2\left(\Lambda_{eff}^A-2\epsilon m^2\right)>0,\quad\quad\quad
\Lambda_{eff}^A\le M_{gr}^2(A)=-\epsilon m^2+\frac{1}{2}\Lambda_{eff}^A\label{bht}
\end{eqnarray}
to take place. They  impose  further restrictions on the values of the cosmological constant $\Lambda_{eff}^{a}=-\kappa ^2W_a=-\frac{1}{L_a^2}$ of type (a) critical points (i.e. on NMG vacua) :
\begin{eqnarray}
0\le\frac{\kappa^2W_{a}^2}{2\epsilon m^2}\le2, \quad\quad \epsilon m^2>0\label{bhta}
\end{eqnarray}
and consequently on  the central charges (\ref{cc}) of the corresponding CFT's. The type (b) NMG vacuum is known to be always unitary \cite{nmg} and whether it represents  UV or IR critical point of the dual $pCFT_2$  depends on the sign factor only : UV - for $\epsilon=-1$ since we have $y_b>0$ and IR - for  $\epsilon=1$ case. The properties of the type (a) critical points ( UV or IR ) do depend on both - the sign of $\epsilon$ and on  the particular form of the matter superpotential, as one can see from eq. (\ref{sdim}).
 
\setcounter{equation}{0}   
\section{Quadratic Superpotential models}
 \label{quadratic}
Let us consider the vacuum structure and related $CFT_2$ data of  NMG model (\ref{acao}) with quadratic superpotential 
\be
W(\s) =  B \s^2 + D ,\;\;\;\;\;\;\ D\neq 0
\ee
 introduced  in ref.\cite{nmg} , where its DW's solutions have been  found. It represents the simplest  example of extended 3D gravity, whose  holographically dual $pCFT_2$ model still permits rather explicit description and as we shall see it exhibits a rich spectrum of different critical phenomena. Its $\beta-$function (\ref{rg}) is parametrized by five parameters (B, D, $m^2$,  $\epsilon$, $\kappa^2=16\pi l_{pl}$) - the same that determine the shape of the matter potential $V(\sigma)$ according to eq.(\ref{sis}). It is important to remember that the classification of the qualitatively different solutions of the RG eq. (\ref{rg}) that describe different critical behaviours of the corresponding 2D dual models requires the complete specification of the qualitatively different regions of the above mentioned parameter space, namely the number and the type of the RG critical points in function of the values of the superpotential's parameters. Independently on the values of the parameters $B$
 and $D$ we always  have  one type $(a)$ vacuum $\sigma_{a}^{*}=0$ represented by $AdS_3$ of cosmological constant $\Lambda_{eff}^{a}(\sigma_{a}^{*})=-\kappa^2D^2$. The $CFT_2(a)$ describing this critical point has central charge given by eq.(\ref{cc}) with $L_{a}^2=\frac{1}{\kappa^2D^2}$. 
 
 
We choose to further investigate the particular case of $\epsilon m^2>0$ only, where we can have in principle few type (b) critical points and fix the sign of $B>0$. Then the available  type $(b)$ RG fixed points, determined by the real roots of equation $W^2(\sigma_{b}^*)=\frac{2\epsilon m^2}{\kappa^2}$ are given by:
\begin{eqnarray}
(\sigma_{\pm}^{*})^2=\pm\frac{\sqrt{2\epsilon m^2}}{\kappa B}-\frac{D}{B}, \quad\quad (\sigma_{-}^{*})^2\le(\sigma_{+}^{*})^2
\end{eqnarray}
Note that there exist two \emph{critical values} of D:
\begin{eqnarray}
D_{cr}^{\pm}=\pm D_{cr},\quad\quad\quad D_{cr}=\frac{\sqrt{2\epsilon m^2}}{\kappa}=\frac{1}{\kappa L_{gr}}\label{Dcr}
 \end{eqnarray}
for which two of the  (b) vacua : $\pm |\sigma_{+}^*|$ or $\pm |\sigma_{-}^*|$ coincide with the (a) one $\sigma_a=0$, giving rise to an inflection (i.e.massless) point $V^{''}(\sigma_a)=0$ of the matter potential. It is then clear that depending on the values of $D$ we have to distinguish  the following three regions in the parameters space : 
\begin{itemize}
\item Region (1): \emph{no} type $(b)$ vacuum for $D>D_{cr}$; 
\item Region (2): \emph{two} type $(b)$ vacua $\{\pm |\sigma_{+}^*\}$ for $-D_{cr}<D<D_{cr}$;
\item Region (3):  \emph{four} type $(b)$ vacua $\{\pm |\sigma_{+}^*|,\pm |\sigma_{-}^*|\}$ for  $D<-D_{cr}$.
\end{itemize}
Remembering  the definitions of the two NMG scales $L_a^2=\frac{1}{\kappa^2 D^2}$ and $L_{gr}^2$, we realize  that the above division of the parameters space in regions of different number of critical points  is  in fact determined by the relations between these scales: $L_a>L_{gr}$ for region (2); $L_a=L_a^{cr}=L_{gr}$ on the borders  (2)-(1) and (2)-(3); $L_a<L_{gr}$ for the both regions (1)and (3).

The  corresponding critical exponents $y_A$, defining the scaling dimension $\Delta_A=2-y_A$ of $\Phi_{\sigma}^{(A)}$, can be obtained from eqs. (\ref{sdim}) for $D\ne 0$\footnote{We exclude here the particular case $D=0$, that corresponds to a flat $E_3$ NMG vacuum, i.e. $\Lambda_a(D=0)= 0$, and $CFT_2$ of $c_{a}(D=0)=\infty$.} :  
\begin{eqnarray}
y_a=\frac{4\epsilon BL_a}{\kappa}\left(1-\frac{ L^2_{gr}}{L^2_a}\right),\quad\quad\quad
y^{\pm}_b=\frac{16\epsilon B L_{gr}}{\kappa}\left(\frac{L_{gr}}{L_a} \mp 1\right)\nonumber
\end{eqnarray}

At $D=\pm D_{cr}$, i.e. on the borders between the regions we have  $y_a(\pm D_{cr})=0$ and therefore  such critical point describes  an infinite order phase transition, as we have mentioned  in section \ref{cftdata}. The vacua of type (a) and (b) together with the two singular points $\sigma_{s}^2 =-\frac{D}{B}$ (that exist for $D<0$ only) divide the  coupling space  $\sigma \in R$ in few intervals to be recognized as different phases of  QFT$_2$ model. Due to the $Z_2$ symmetry $\sigma\rightarrow -\sigma$ of the NMG model with quadratic superpotential the phase structure in all of the regions remains invariant under $\sigma$ reflections, i.e. it is  enough to study only the phases corresponding to $\sigma\ge 0$. For example, in the \emph{region} (2) we have to consider separately the case $(2+)$ for $D\in (0,D_{cr})$ of two critical points only : $(0_{IR },|\sigma_{+}^{*}|_{UV}) , \quad (|\sigma_{+}^{*}|_{UV},\infty)$ and exhibiting only two phases;  from the $(2-)$ one for $D\in (-D_{cr},0)$ , where due to the presence of two singular points we find the following three phases: $(0_{UV},|\sigma_s|),\quad(|\sigma_s|,|\sigma_{+}^{*}|_{UV}), \quad (|\sigma_{+}^{*}|_{UV},\infty)$. We note that the index UV or IR above ( say $0_{IR}$ ) marks  the type of the RG fixed points (related to the sign of $y_A$) for the case $\epsilon=-1$ and $m^2<0$. In the case of $\epsilon=1$ and $m^2>0$ the corresponding $(2\pm)-$phase structure is identical to the above one, but now with $UV$ and $IR$ \emph{interchanged}. The \emph{region} (3) has the richest phase structure formed by the following four phases:
\begin{eqnarray}
(3)\quad \epsilon=-1 :\quad (0_{IR},|\sigma_{-}^{*}|_{UV})\quad(|\sigma_{-}^{*}|_{UV},|\sigma_s|) \quad
(|\sigma_s|,|\sigma_{+}^{*}|_{UV})\quad (|\sigma_{+}^{*}|_{UV},\infty).\label{re3}
\end{eqnarray}
while the  \emph{region} (1) - the simplest one: $(0_{UV},\infty)$ only.

Few comments about the dependence of the properties of the solutions of the RG eq.(\ref{rg}) on the values of the parameter B are in order. Let us remind that the above described ``phase" structure of the parameters space was derived under the condition $B>0$ and varying the remaining parameter D of the superpotential. It is straightforward to verify that the corresponding results for \emph{negative} values of B can be obtained from the ones of \emph{positive} values of B (as above) by applying  the following formal  rules
\begin{eqnarray}
B \rightarrow -B : \ \sigma_{+}^* \rightarrow \sigma_{-}^*,\ reg.(3)\rightarrow reg.(1),\ reg.(2+)\rightarrow reg.(2-),\ \epsilon \rightarrow -\epsilon,\label{B}
\end{eqnarray}
and \emph{without} changing the index UV or IR of the critical points.

\setcounter{equation}{0}
\section{On the properties of the  DW's solutions}
\label{properties}

Consider an arbitrary domain wall solution DW$_{-+}$ of eqs. (\ref{sis}) relating two consecutive vacua $(\sigma_{\pm},L_{\pm},y_{\pm})$, with $\sigma_{-}<\sigma_{+}$. It has the form (\ref{intro1}), with the following b.c.'s at $z \rightarrow\pm \infty$:
\begin{eqnarray}
\sigma(\pm \infty)=\sigma_{\pm}\quad\quad \dot{\varphi}_{\pm}=\dot{\varphi}(\pm\infty)=-2\epsilon\kappa W_{\pm},\quad\quad W(\sigma_{\pm})=W_{\pm}.   \label{bcs}
\end{eqnarray}
Depending on the specific shape of the NMG superpotential, we can have a few different combinations of couples of b.c.'s defining qualitatively different DW's geometries \cite{nmg} . We next consider the definitions of the \emph{four} types of admissible DW's, and present some simple representative examples for most of them.    

\emph{1. Standard boundary/horizon AdS$_3$/AdS$_3$ DW's.} Let us first consider a superpotential $W(\sigma)$ which has no zeros for $\sigma \in (\sigma_{-},\sigma_{+})$, thus both $W_{\pm}$ have the same sign and we also assume that $\sigma_{+}$ is a critical point of UV -type, i.e. $0<y_{+}<2$. Then we must take $\epsilon\kappa \, W_{\pm}= - 1/L_{\pm}$  in order to ensure that the ``UV-vacuum" $(\sigma_{+},L_{+},y_{+})$ represents, asymptotically, an AdS$_3$ boundary geometry, and the ``deep bulk IR-region " (i.e. the space around the IR vacuum $(\sigma_{-},L_{-},y_{-})$), of vanishing scale factor, corresponds to a null Cauchy horizon\footnote{In the euclidean case, the corresponding $(a)H^3$ geometry is given by $t \rightarrow i\tau$. Then this limit is just one point that should be added to the $2-d$ boundary to complete the 2-sphere $S^2$ representing the conformal compactification of the 2$d$-euclidean plane $E_2$.}:
\begin{eqnarray}
e^{\varphi}\stackrel{z\rightarrow\infty}{\approx}e^{-2\epsilon\kappa W_{+}z}\rightarrow\infty,\quad\quad
e^{\varphi}\stackrel{z\rightarrow-\infty}{\approx}e^{-2\epsilon\kappa W_{-}z}\rightarrow 0 .\nonumber
\end{eqnarray}

\emph{2. Janus two-boundaries $AdS_3/AdS_3$ DW's.} In the case when $W(\sigma)$ has one zero $\sigma_s\in (\sigma_{-},\sigma_{+})$, and consequently $W_{\pm}$ have opposite signs, we must introduce another identification  $\epsilon\kappa W_{\pm}=\mp 1/L_{\pm}$. It is then evident from eqs. (\ref{bcs}) that the scale factor is now divergent (or zero) at both vacua, which are both of the same UV or IR type. Such DW's define a particular $(a)AdS_3$ geometry of Janus type 
\cite{freed}\,\cite{nmg} presenting, for $\epsilon=-1$, \emph{two different boundaries} (or two horizons, for $\epsilon=1$). The corresponding geometries are rather different from the standard $(a)AdS_3$ ones with \emph{one boundary and one horizon}, introduced above (see refs. \cite {nmg}\,\cite{pos} for more details). The simplest explicit example of a Janus DW is provided by the linear superpotential $W=A\sigma$ kink-like solutions, with $\epsilon m^2 < 0$, $A^2 < |m^2|$ and $\sigma_{\pm} = \pm 1/\kappa L_{gr}A$, $\sigma_s = 0$:
\begin{eqnarray}
&& \sigma(z)=\frac{1}{\kappa L_{gr}A}\tanh\Big(2A^2L_{gr}(z-z_{0})\Big), \nonumber \\
&& e^{\varphi(z)-\varphi_{0}}=\Big[\cosh\Big(2A^2L_{gr}(z-z_{0})\Big)\Big]^{\frac{4}{y_{+}}}, \;\;  y_+ = y_- = 4A^2 L^2_{gr} .    \label{jdw}
\end{eqnarray}
They have as asymptotics at $z\rightarrow\pm\infty$ two very special NMG \emph{unitary} vacua with $\lambda=- \Lambda/ m^2 = -1$, representing two AdS$_3$ spaces of \emph{equal} cosmological constants $\Lambda^{\pm}_{\text{eff}}=- 1/L^2_{gr}$. Note that $L_{+}=L_{-}$ is a particular property of the Janus DW's relating two UV type (b) vacua, while for the corresponding  Janus DW's between one type (a) and one type (b) vacua, we have indeed $L_{-}\neq L_{+}$. In both cases the scale factor reaches its (finite) minimum at the point $\sigma_s=0$.
  
 \emph{3. AdS$_3$/flat DW's.} Within the family of the standard $AdS_{3}(L_{-})/AdS_{3}(L_{+})$  ``boundary/horizon" type of DW's, we have to separately consider the limiting case of say $L_{-}\rightarrow \infty$ (i.e. $\Lambda^{\text{eff}}_{-}\rightarrow 0$), corresponding to  the flat $E_3$ (or $M_{2,1}$) vacuum, cf. eqs.(\ref{intro1}) and (\ref{eq8}), with $e^{\varphi}\stackrel{z\rightarrow-\infty}\rightarrow {\mathrm{constant}} \neq 0$. These vacua do not correspond to critical points at all. As we have shown (cf. eq.(\ref{rg})), the $\beta$-function is divergent at such points and furthermore the E$_3$ metric is not scale invariant, since its isometry group --- the 3D Poincar\'e group --- is different from the conformal one, $SO(3,1)$, that describes the isometries of the euclidean AdS$_3$ space. In fact, they have a rather different 2D statistical mechanics or/and QFT$_2$ interpretation. It turns out that such ``flat" vacua play an important role in the description of certain \emph{massive} RG flows that occur in the dual pCFT$_2$. The particular DW's of AdS$_3$/E$_3$ type (interpolating between one AdS$_3$ vacuum of UV type and one such E$_3$ vacuum in the deep IR) represent the NMG geometrical counterpart of specific massive phases in the corresponding 2D model%
\footnote{To be compared with the AdS$_3$/AdS$_3$ type of DW's \cite{nmg} relating two different AdS$_3$ vacua, corresponding both to critical points (i.e. to certain CFT$_2$'s), and thus describing a \emph{massless} RG flow.}.
 In order to make transparent the properties of this kind of flat vacua, we present here the explicit DW solution for the quadratic superpotential $W=B\sigma^2+D$ with $D=0$ (we fix $B>0$ and $\epsilon m^2<0$). Within the interval $\sigma \in (0,\sigma_{+})$, it has a flat vacuum at $\sigma_{-}=0$ with $W_{-}=W(0)=0$: 
\begin{eqnarray}
\sigma(z)=\frac{1}{\sqrt{\kappa L_{gr}B}} \Big(1+e^{-\frac{16B}{\kappa}(z-z_{0})}\Big)^{-1/4},\quad
e^{\varphi(z)-\varphi_{0}}=\Bigg(\frac{1-\kappa L_{gr}B\sigma^{2}(z)}{1+\kappa L_{gr}B\sigma^{2}(z)}\Bigg)^{\kappa\epsilon/8BL_{gr}} .   \label{flat}
\end{eqnarray}
This example represents an interesting self-dual pCFT$_2$ model with two massive phases, whose features require further investigations.

\emph{4. AdS$_3$/n.s. singular DW's.} Let us complete the list of the qualitatively different DW's solutions of the NMG-matter model (\ref{acao}) by adding the case of ``singular" DW's relating one UV-type vacuum  with a naked singularity (n.s.) at $z = z_0$, i.e. $R^{(3)}(z_{0})=-\infty$. They are defined in the interval $z\in(z_{0},\infty)$ and correspond to the following specific b.c.'s: $\sigma(z\rightarrow \infty)=\sigma_{+}$, $\sigma(z_{0})=\infty$ and  $e^{\varphi}\stackrel{z\rightarrow z_0}\rightarrow {\mathrm{constant}} \neq 0$. An example of such DW's is the solution for quadratic superpotential with $D=0$, but now within the interval $\sigma \in (\sigma_{+},\infty)$:
$$\sigma(z)=\frac{1}{\sqrt{\kappa L_{gr}B}} \Big(1-e^{-\frac{16B}{\kappa}(z-z_{0})}\Big)^{-1/4} . $$

Let us mention the important fact that both DW's for the quadratic $D=0$ superpotential -- flat/$AdS_3(L_{+})$ and $AdS_3(L_{+})$/n.s. -- we have considered above share a common boundary at $z\rightarrow\infty$ given by the UV vacua $\sigma_{+}$. This is quite a general property, valid for a generic superpotential with various different NMG vacua $\sigma_{a_k},\sigma_{b_k}$ and zeros at $\sigma_s$  ($k=1,2,...N$) which divide the entire coupling space $\sigma\in \mathbb{R} $ into intervals $p_{k,k+1}=(\sigma_{A_k},\sigma_{A_{k+1}})$ of a few different kinds, each one corresponding to a DW of one the above considered four types: (1) n.s./AdS$_3$ or AdS$_3$/n.s.; (2) $AdS_3/AdS_3$ of Janus two boundaries; (3) $AdS_3/AdS_3$ of standard boundary/horizon type 
and (4) flat/AdS$_3$ or AdS$_3$/flat type.  Notice that the change of the IR-type of b.c.'s that occurs at each one of the boundaries describes in fact  the \emph{transition} between two different types of $(a)AdS_3$ geometries depending on the interval $p_{k,k+1}=(\sigma_{A_k},\sigma_{A_{k+1}})$ to which the initial ``RG" condition $\sigma(l=0)=\sigma_0$ belongs to.

\setcounter{equation}{0}
\section{Holographic RG flows and phase transitions} 
{\label{holoflows}

The $CFT_2$'s data $(\sigma_n^*,c_n,y_n)_{UV/IR}$ specific for each parameters space region, provide  the boundary conditions necessary for the derivation of the solutions of RG eqs. (\ref{rg}) and (\ref{fs}) characterizing each phase $p_{nk}=(\sigma_n^*,\sigma_k^*)$. The RG flows by definition represent the way the coupling constant $\sigma(l,D)$ is running between two neighbour critical points  when the RG scale $L_*$ increases from $L_*^{UV}=0$ (i.e. $l_{UV}=\infty$) to $L_*^{IR}=\infty$ (i.e. $l_{IR}=-\infty$). Depending on the behaviour of the correlation length $\xi(\sigma)$, the s.p. of the free energy $F_s(\sigma)$ (and its derivatives) and of the correlation functions $G_{\Phi}(x_{12},\sigma)$ we distinguish in the non-degenerate case $D\ne\pm D_{cr}$  the following three types of phases:
\begin{eqnarray}
&(1)& massless \ (UV/IR):0<L_*\le\infty \quad \xi(\sigma_{UV}^*)\approx \infty, \xi(\sigma_{IR}^*)\approx 0; \ \sigma(-\infty)=\sigma_{UV}^*,\sigma(\infty)=\sigma_{IR}^*,\nonumber\\    
&(2)&massive  \ (UV/\infty) : 0<L_*\le L_*^{ms}\quad \xi(\sigma_{UV}^*)\approx \infty,\quad \xi(\sigma\approx\infty)\approx L_*^{ms},\nonumber\\ 
&(3)&Janus \  (UV_{+}/\sigma_s/UV_{-}):0<L_*\le L_*^{max}\quad \xi(\sigma_{UV}^{\pm})\approx \infty,\quad \xi(\sigma_s)\approx L_*^{max}\label{phase}
\end{eqnarray}
\vspace{0.5cm}


The simplest example is provided by the phase structure of $pCFT_2$ model in region $(2+)$ \cite{nmg} . For $\sigma>0$ and $\epsilon=-1$ it contains two phases: $p_{ml}=(0_{R},|\sigma_+^*|_{UV})$ and $ p_{ms}=(|\sigma_+^*|_{UV},\infty)$, characterized by the singularities and asymptotic behaviour of the solutions of eqs. (\ref{rg}), ({\ref{fs}):  
 \begin{eqnarray}
\xi_{(2+)}(\sigma,\sigma_0)\approx e^{-l} =\left(\frac{\sigma^2}{\sigma_0^2}\right)^{-\frac{1}{2y_0}}
\left(\frac{(\sigma_+^*)^2-\sigma^2}{(\sigma_+^*)^2-\sigma_0^2}\right)^{-\frac{1}{y_+}}\left(\frac{(|\sigma_-^*)|^2+\sigma^2}{|(\sigma_-^*)|^2+\sigma_0^2}\right)^{-\frac{1}{y_{-}}}, \label{ft1}
\end{eqnarray}
where  $\sigma_0=\sigma(l=0)$ represents the ``initial" condition of RG rescalings, i.e. $L_*^{(0)}\approx 1$ . The critical behaviour of the s.p. of the free energy $F^{(2+)}_s(\sigma)=e^{2l}$ and the properties of the correlation length within $p_{ml}=(0_{IR
},|\sigma_+^*|_{UV})$ -- starting at the UV critical point as $\xi_{(2+)}(\sigma_+^*|_{UV},\sigma_0)\approx \infty$  and terminating at the the IR on $0_{IR}$, where we have $\xi_{(2+)}(0_{IR},\sigma_0)\approx 0$ -- allow us to identify 
$(0_{IR},|\sigma_+^*|_{UV})$ as a massless phase. In this case (i.e. for $\epsilon=-1$ and $B>0$, $L_a > L_{gr}$) the \emph{massless} RG flow is driven by the (relevant)\footnote{we are further requiring $B <\frac{\kappa L_a}{8L_{gr}(L_a-L_{gr})}$ in order to ensure that $0<y_{UV}=y_{b+}<2$, which is the condition for relevance $0<\Delta^{UV}<2$ of this operator} operator $\Phi^{UV}_{\sigma}$ of dimension $\Delta^{UV}_{\Phi}=2-16\epsilon B L_{gr}(L_{gr}-L_a)/\kappa L_a<2$ from the UV-$CFT_2$ of central charge $c_{UV}=3L_{gr}/l_{pl}$. According to the discussion in section \ref{cftdata} (see also \cite{nmg}) the corresponding IR-$CFT_2$ has central charge $c_{IR}=\frac{3\epsilon L_a}{2l_{pl}}\left(1+L_{gr}^2/L_a^2\right)$ and due to the renormalization of the coupling constant during the flow, the \emph{renormalized} operator $\Phi_{\sigma}$ becomes irrelevant at the IR limit of dimension $\Delta^{IR}=2-y_{a}=2-4\epsilon BL_a(1- L^2_{gr}/L^2_a)/\kappa >2$. We have to note that although for $\epsilon=-1$ both the central charges turn out to be negative, one can easily verify that 
$c(|\sigma_+^*|_{UV})> c(0_{IR})$, i.e.  the central charge is  decreasing during the  massless flow as required from the c-theorem \cite{x} . In the case ($\epsilon=1,m^2>0$) the direction of the flow is inverted, since now we have that $y(0)=y_a>0$ and $0_{IR}$ becomes a critical point of UV type, while  $y(|\sigma_+^*|)<0$ and therefore $|\sigma_+^*|_{UV}$ is of IR-type. Nevertheless, the central charge is again decreasing during the flow, since now both the central charges are positive and we realize that indeed 
$c(0_{IR})>c(|\sigma_+^*|_{UV})$. Although we have no characteristic (mass) scale in this interval $\sigma \in p_{ml}=(0_{R},|\sigma_+^*|_{UV})$, the considered $pCFT_2$ model is \emph{not} conformal invariant.

The  $(2+)$-phase corresponding to the coupling space interval $ \sigma\in  p_{ms}=(|\sigma_+^*|_{UV},\infty)$  is characterized by the \emph{finite} value of correlation length for $\sigma \rightarrow\infty$: 
 \begin{eqnarray}
&&\xi_{(2+)}(\sigma \rightarrow\infty, \sigma_0>|\sigma_+^*|_{UV})\approx e^{-l_{ms}}\nonumber\\
&&= \left(\sigma_0^2\right)^{\frac{1}{2y_0}}
\left(\sigma_0^2-(\sigma_+^*)^2\right)^{\frac{1}{y_+}}\left(|(\sigma_-^*)|^2+\sigma_0^2\right)^{\frac{1}{y_{-}}} \label{ft2}
\end{eqnarray}
 as one can easily verify  from the limit of eq.(\ref{ft1}) taking into account the remarkable
 "resonance" property $\frac{1}{2y_0}+\frac{1}{y_+}+\frac{1}{y_-}=0$, specific for our quadratic superpotential\footnote{Which  in fact assures that the ``naked singularity" $\sigma=\infty$ is \emph{not} a critical point.}. We observe that in  this phase  the coupling constant runs to infinity while the RG scale is running in the finite interval $L_*\in (0,L_*^{(ms)})$ thus defining particular \emph{mass gap} 
 \begin{eqnarray} 
   M_{(ms)}\approx \frac{1}{L_*^{(ms)}}=\left(\sigma_0^2\right)^{-\frac{1}{2y_0}}
\left(\sigma_0^2-(\sigma_+^*)^2\right)^{-\frac{1}{y_+}}\left(|(\sigma_-^*)|^2+\sigma_0^2\right)^{-\frac{1}{y_{-}}}\label{mass1}
\end{eqnarray}
 As a consequence the corresponding $\Phi_{\sigma}$ correlation function (\ref{fs}) changes its behaviour including now at the  leading order  specific exponential decay term $e^{-M_{ms}|x_{12}|}$ that determines the massive properties of this  $pCFT_2$-phase. We have therefore  an example of phase transition from massless to the  massive phase that occurs at the UV critical point $|\sigma_+^*|$ in the $(2+)$-phase of $pCFT_2$ model. The 3d gravity description of such phase transition involves \emph{two different} NMG solutions having coinciding boundary conditions $(|\sigma_+^*|,\Lambda_{eff}^+, \Delta_+)$ at their common boundary $z\rightarrow\infty$, i.e. at $\sigma(\infty)=|\sigma_+^*|$. The massive phase is ``holographically" described by \emph{singular} DW  metrics giving rise to $(a)AdS_3$ space-time with naked singularity \cite{nmg} , while the massless one corresponds to the regular DW (constructed in ref.\cite{nmg}) interpolating between the two NMG vacua $|\sigma_+^*|_{UV}$ and $0_{R}$.
 
 The above analysis of the critical phenomena in $pCFT_2$ model (and their 3D geometrical counterparts) based on the standard statistical mechanical and RG methods, allows us to establish the \emph{basic rule} of the off-critical  $(a)AdS_3/CFT_2$ correspondence, namely: the NMG-geometrical  description of the \emph{phase transitions} in its dual $pCFT_2$ model is given by the analytic properties - poles, zeros, cuts and essential singularities - of the\emph{ scale factor} $e^{\varphi}$ of 3D DW's metrics of the $(a)H_3$ (euclidean) type:
\begin{eqnarray}
F_s^{(2+)}(\sigma,\sigma_0)\approx e^{2l} \approx\xi_{(2+)}^{-2}\approx e^{-\varphi(\sigma)} \label{rule}
\end{eqnarray} 
as a function of the matter field $\sigma$ obtained by excluding the radial variable $z$ from the corresponding DW's solutions \cite{nmg} . Another important ingredient of the \emph{off-critical holography} is the so called Zamolodchikov's central function for NMG model \footnote{It represents a natural generalization \cite{sinha} of the well known result for $m^2\rightarrow\infty$ limit \cite{4}, \cite{VVB}} introduced in refs. \cite{sinha}\,\cite{nmg} :
\begin{eqnarray}
C(\sigma)=-\frac{3}{2G\kappa W(\sigma)}\bigg(1+\frac{\kappa^2W^2(\sigma)}{2\epsilon m^2}\bigg)\label{cf}
\end{eqnarray}
which at the critical points $\sigma_{A_{\pm}}^*$ takes the values (\ref{cc}). Remember that according to the $I^{st}$ order eqs.(\ref{sis}) we have $W(\sigma)=-\frac{\dot\varphi}{2\epsilon\kappa}$ and therefore the central charges $c_A$ and the central function as well  are geometrically described by the log-derivative $\dot\varphi$ of the scale factor. As a consequence of its definition (\ref{cf}) and of  the RG eqs. (\ref{rg}) we  conclude that \cite{nmg} :
\begin{eqnarray}
 \frac{dC(\sigma)}{dl} = -\frac{3}{4GW(\sigma)}\left(\frac{d \sigma}{dl}\right)^2 \label{cf1}
\end{eqnarray}
Hence for $W(\sigma)$ positive (as in our example) the central function is decreasing during the  massless flow, i.e. we have $c(|\sigma_+^*|_{UV})> c(0_{IR})$ for $\epsilon=-1$. 

Few comments concerning the identification of this NMG induced holographic phase structure with the one of certain perturbed $CFT_2$ model are in order. All the properties of the massless-to-massive phase transition observed in the dual QFT$_2$ in the region $(2+)$ obtained from the NMG-induced exact $\beta$-function turns out to be almost identical to those calculated perturbatively in pCFT$_2$, with action (\ref{eq28}), where the $CFT_2$ in the UV critical point and the corresponding relevant operator are chosen to coincide with those we have found above. It is important to also mention that the operator $\Phi^{UV}_{\sigma}$ of dimension $\Delta^{UV}_{\Phi}$ has OPE of $\Phi_{adj}$ (or $\Phi_{13}$) type \cite{bpz} with  structure constant given by
$C_{\Phi\Phi\Phi}(|\sigma_{+}^{*}|_{UV})=8\epsilon B^2 L^2_{gr}|\sigma_{+}|(1+2\frac{L_{gr}}{L_a})$, similar to the conformal OPE's we have in the Virasoro  and Liouville  models \cite{x}\,\cite{fat} . In order to make this equivalence exact one need to  identify  the NMG -matter model parameters $L_{gr},L_a$  with the one (denoted usually by $b$) of the corresponding $CFT_2$'s central charges $c_{\pm}(b) =  1 \pm 6 Q_b^2$ with $Q_b = b \pm \frac{1}{b}$}. Further investigations are  needed  in order to find an appropriate large $c^{UV}$ (i.e. large $L_{gr}/l_{pl}$) limits consistent with the conformal perturbation theory.

 The RG flows in region $(2-)$ are rather different from the ones of (2+) due to the fact that all the critical points are now of UV-type  and to the presence of singular points as well. The massive phase $(|\sigma_{+}^{*}|_{UV},\infty)$ coincides with the corresponding one in reg.(2+) and the mass gap is given again by $M_{ms}$ of eq. (\ref{mass1}) except that the values of the exponents $y_{\pm}>0$ and $y_0>0$ are different due to negative sign of $D<0$ in this region. The new massive phase is related to the Janus type DW's solutions (see sect.4. above)  connecting two critical points (NMG vacua) with singular point in between, i.e. $0_{UV}/|\sigma_s|/|\sigma_{+}^{*}|_{UV}$  both provided with \emph{relevant} operators. As one can see from  the scale factor and from the correlation length $\xi_{(2-)}$ behaviours (\ref{ft2}) the RG scale is now start running  from $L_*=0$ at the both $ 0_{UV}$ and $|\sigma_{+}^{*}|_{UV}$ critical points and it gets  its maximal value $L_*^{(max)}$ at the ``end point" $|\sigma_s|=\sqrt{\frac{|D|}{B}}$. Note the important difference with the normal massive phase where the $L_*^{(ms)}$ was reached for $\sigma\approx \infty$. The proper existence of $L_*^{(max)}$ however introduces  mass scale:
 \begin{eqnarray}
M_J^{(2-)}(\sigma_s,\sigma_0)\approx e^{l^{max}} =\left(\frac{\sigma_s^2}{\sigma_0^2}\right)^{\frac{1}{2y_0}}
\left(\frac{(\sigma_+^*)^2-\sigma_s^2}{(\sigma_+^*)^2-\sigma_0^2}\right)^{\frac{1}{y_+}}\left(\frac{(|\sigma_-^*)|^2+\sigma_s^2}{|(\sigma_-^*)|^2+\sigma_0^2}\right)^{\frac{1}{y_{-}}}
\end{eqnarray}
specific for the new Janus-massive phase. Hence in this case we have two different  massive phases that start from the same critical point  $|\sigma_{+}^{*}|_{UV}$. This massive-to-massive phase transition is characterized by the ratio of the two mass gaps:     
\begin{eqnarray}
\frac{M_J^{(2-)}}{M_{(ms)}^{(2-)}} = \left(\frac{L_a}{L_{gr}}\right)^{\frac{1}{y_{+}}}
\left(2-\frac{L_a}{L_{gr}}\right)^{\frac{1}{y_{-}}},\label{mrat}
\end{eqnarray} 
which differently from the corresponding $\xi$'s and mass gaps is completely determined by the superpotential data and turns out to be an important characteristics of corresponding QFT$_2$ model. The NMG description of $(2-)-$ phase diagram is therefore given  by one Janus-type DW and one singular solution representing naked-singularity.

The phase structure of the $pCFT_2$ model 
 corresponding to region (3) turns out to combine all the critical  phenomena  observed in regions $(2\pm)$. Consider again the $\sigma>0$ case. The coupling space is now divided in four intervals (i.e.phases): 
 $$(0_{IR},|\sigma_{-}^{*}|_{UV}),\quad(|\sigma_{-}^{*}|_{UV},|\sigma_s|),\quad
(|\sigma_s|,|\sigma_{+}^{*}|_{UV}), \quad (|\sigma_{+}^{*}|_{UV},\infty)$$ 
containing three critical  and one singular points. As one can verify from the behaviour of the corresponding correlation length: 
\begin{eqnarray}
\xi_{(3)}(\sigma,\sigma_0)\approx \left(\frac{\sigma^2}{\sigma_0^2}\right)^{-\frac{1}{2y_0}}
\left(\frac{(\sigma_+^*)^2-\sigma^2}{(\sigma_+^*)^2-\sigma_0^2}\right)^{-\frac{1}{y_+}}\left(\frac{(|\sigma_-^*)|^2-\sigma^2}{|(\sigma_-^*)|^2-\sigma_0^2}\right)^{-\frac{1}{y_{-}}} \label{ft2}
\end{eqnarray} 
the phase $(0_{IR},|\sigma_{-}^{*}|_{UV})$ is describing massless RG flow  similar to the one in the region $(2+)$ but involving the new critical point $|\sigma_{-}^{*}|_{UV}$. The next two phases $(|\sigma_{-}^{*}|_{UV},|\sigma_s|)$ and
$(|\sigma_s|,|\sigma_{+}^{*}|_{UV})$ are both representing Janus-massive phases, while the last one $(|\sigma_{+}^{*}|_{UV},\infty)$ is identical to the mass phase of region $(2+)$ except that the exponents $y_A$ with $A=\pm,0$ as well as the mass gap formula are the ones specific for the region (3) with $D<-D_{cr}$, i.e. $L_a<L_{gr}$. It is evident that the holographic description of the reg. (3) phase structure in terms of NMG's DW solutions consists of three different DW's of common boundaries: one of UV-IR type interpolating between two different NMG vacua of cosmological constants $\Lambda_0^{eff}$ and $\Lambda_{-}^{eff}$, one of Janus type and the last one that involves  naked singularity as defined in Sect.4 above.

The nature of the phase transitions in QFT$_2$ occurring  at the borders of the parameter space $D=\pm D_{cr}$ (i.e. for $ L_a=L_{gr}$) where $y_a(\pm D_{cr})=0$  is rather different from the ones of $II^{nd}$ order we have described above. As one can see by comparing the  forms of the  corresponding solutions $\sigma=\sigma(l,D)$ of RG equations (\ref{rg}), (\ref{fs}):
 \begin{eqnarray}
&&\xi(D=\pm D_{cr})\approx e^{-l}\approx e^{\frac{\varphi}{2}} \nonumber\\
&&=\left(1\pm\frac{2\sqrt{2\epsilon m^2}}{\kappa B\sigma^2}\right)^{\frac{1}{y_{\mp}(cr)}} e^{\frac{\rho_0}{\sigma^2}-\frac{2}{y_{\mp(cr)}}},
\quad y_{\mp}(cr)=\pm\frac{32\epsilon B}{\kappa \sqrt{2\epsilon m^2}}\nonumber\\
&&\xi(D\ne\pm D_{cr})\approx e^{-l}\approx e^{\frac{\varphi}{2}} \nonumber\\
&&=\left(\sigma^2\right)^{-\frac{1}{2y_0}}
\left((\sigma_+^*)^2-\sigma^2\right)^{-\frac{1}{y_+}}\left((\sigma_-^*)^2-\sigma^2\right)^{-\frac{1}{y_{-}}},\rho_0 =-\frac{m^2}{8B^2} \label{xi}
\end{eqnarray}
the specific power-like singularities for $D\ne\pm D_{cr}$ are replaced now by an essential singularity at the (here triple) zero of the $\beta-$function. According to definitions of section \ref{cftdata},
 the  presence of such singularity  in $\xi(\sigma,D_{cr})$ and $F_s^{cr}(\sigma)$ at the critical point $\sigma= 0_{mar}$ in the case $D=D_{cr}$ is an indication of infinite order phase transition. The corresponding massive phase $(0_{mar},\infty)$  is characterized by the mass gap $M_{ms}^{cr}\approx e^{2/y_{\mp(cr)}}$ obtained from  the $\sigma\rightarrow \infty$ limit of eq. (\ref{xi}). The phase structure in the case $D=-D_{cr}$ is richer: for $\sigma >0$ we have one "marginal" critical point $0_{mar}$, one singular point $|\sigma_s^{cr}|$ and one UV critical point $|\sigma_{+}^{cr}|_{UV}$  giving rise to four \emph{massive} phases. The first two weak-coupling massive  phases $(0_{mar},|\sigma_s^{cr}|)$
and $(|\sigma_s^{cr}|,|\sigma_{+}^{cr}|_{UV})$ are of Janus type with  mass gap  given by $M_{J}^{cr}\approx e^{4/y_{\mp(cr)}} =(M_{ms}^{cr})^2$  and the last one $(|\sigma_{+}^{cr}|_{UV},\infty)$ is the standard strong coupling massive phase $(|\sigma_{+}^{cr}|_{UV},\infty)$. The phase transitions at $\sigma= |\sigma_{+}^{cr}|_{UV}$ point is of J-massive-to-massive (weak-to-strong coupling) type, quite similar to  the one observed in the region $(2-)$ above (i.e. of second order), but indeed with  different  mass ratio $\frac{M_{J}^{cr}}{M_{ms}^{cr}}= e^{2/y_{\mp(cr)}}$.

\setcounter{equation}{0}
\section{Concluding remarks}

Our investigation of the \emph{classical} critical phenomena in the $pCFT_2$'s duals to the NMG models with quadratic matter superpotential  has revealed many essential features of these 2D non-conformal models  leaving however still open the problem of their complete identification. It is important to emphasize that  the phase transitions we have described concern the TD limits of certain  2D \emph{classical} statistical models (s.m.), related to the $pCFT_2$ in discussion. We have studied the \emph{infinite volume} critical properties  of these statistical  models by using the Wilson RG methods. As it well known (see section $4.5$ of the Cardy's book \cite{cardy}) the \emph{finite} temperature phase transitions in the classical $d=2$ s.m. in infinite volume correspond to \emph{zero} temperature phase transition in certain equivalent quantum $d=1$ s.m. (or its TD  $1+1$ QFT limit) when  some other coupling in the quantum model becomes critical, say the transverse magnetic field in the case of 1D Ising model. Observe that the temperatures used in both models are different: the inverse temperature (i.e. $1/kT_1$) in the quantum 1D model corresponds to the period of the extra (time) direction, while  the temperature in 2D classical s.m. is related to the extra coupling constant in 1D model. Hence the description of finite $T_1$-phase transitions in the quantum 1D s.m. requires to study the finite-size effects in its 2D classical counterpart. This fact explains the successful use of DW solutions in the descriptions of classical phase transitions (in 2D s.m.) instead of say black holes  and other periodic (or finite time) solutions of NMG model, which are indeed the geometric ingredients required in the investigation of  finite $T_1$ quantum phase transitions.
  
The detailed description of the main features - critical exponents, mass gaps, s.p. of the reduced free energy - of the variety of second and infinite order classical phase transitions in 2D s.m. models that are \emph{conjectured} to be dual of the NMG model ({\ref{acao}), allows us to establish the following important rule of the off-critical $AdS_3/CFT_2$ correspondence: the \emph{phase transitions} observed in the  dual $pCFT_2$ models are  determined by the  analytic properties of the \emph{scale factor} $e^{\varphi}$ of  3D (euclidean) DW's type metrics of NMG model \cite{nmg} written as  a function of the matter field $\sigma$. As we have shown by using RG methods, the inverse of the \emph{scale factor} is proportional of the s.p. of the \emph{free energy}. In order to calculate the exact values of the entropy, the specific heat and other important TD characteristics one need to know the finite part of the free energy as well, which is a rather complicated problem even for the simplest 2D s.m. models. It is well known however that the  infinite 2D conformal symmetry at the \emph{critical points} offers powerful methods, based on the knowledge of the characters of the  Virasoro algebra  representations, which allow to construct the exact form of  the corresponding ``critical" partition functions. 
An important step towards the construction of corresponding \emph{holographic partition functions}  is the calculation of the spectra of the linear fluctuations around the DW's solutions  of the considered NMG models. One has to further consider the difficult problem of the construction of the off-critical correlation functions of 2D fields dual to 3D matter scalar by applying $AdS/CFT$ methods \cite{witt}\, \cite{gub}  and to  next compare with the known results of corresponding 2D models \cite{muss}\,\cite{azz} as well . 

Let us also mention the problem with the interpretation of the negative values of the ''classical''  central charges (\ref{cc}) for $\epsilon=-1$ and $m^2<0$, that are  usually considered as \emph{non-unitary} $CFT_2$'s. According to  the well known exact results for the \emph{quantum} minimal $CFT_2$  \cite{bpz}\,\cite{cardy} , based on the degenerate representations of the Virasoro algebra,  in all the cases when $ c<0$ the corresponding $CFT_2$'s contain primary fields  of negative dimensions  and hence they indeed represent non-unitary $QFT_2$'s. We should remember however also the well established results \cite{fat}\,\cite{zam-zam} concerning  the classical and semi-classical limits $\hbar\rightarrow 0$ of the anomalous dimensions  and of the  the central charges of all the models of the \emph{unitary series} $c_{quant}=1-6\frac{Q^2_{quant}}{\hbar}$. They lead to  large negative central charges $c_{quant}\rightarrow c_{cl}\approx-\infty$. Hence the classical and semi-classical large negative central charges are common feature  for  \emph{both unitary and non-unitary quantum} minimal models of $c_{quant} <1$. Further investigations of the limiting properties (involving the calculation of certain sub-leading terms) of the dimensions of the primary fields an of the structure constants their OPE's are needed in order to conclude whether the strong-coupling CFT's data we have extracted from the NMG holographic dual $ CFT_2$'s are representing limits of  unitary or non-unitary quantum m.m.'s.

\vspace{1 cm}
\textbf{Acknowledgements.} We are grateful to A.L.A.Lima  for critical reading of the manuscript and for pointing out to us  few mistakes and misprints in the e-Print version arXiv:1009.2665 [hep-th], as well as for his suggestions for improvements.

This work has been partially supported by PRONEX project number 35885149/2006 from FAPES-CNPq (Brazil).   











\begin{thebibliography}{99}

\bibitem{malda}
J.~M.~Maldacena,
  ``The Large N limit of superconformal field theories and supergravity,''
  Adv.\ Theor.\ Math.\ Phys.\  {\bf 2} (1998) 231
  [hep-th/9711200].
\bibitem{witt}
 E.~Witten,
  ``Anti-de Sitter space and holography,''
  Adv.\ Theor.\ Math.\ Phys.\  {\bf 2} (1998) 253
  [hep-th/9802150].
 S.~S.~Gubser, I.~R.~Klebanov and A.~M.~Polyakov,
  ``Gauge theory correlators from noncritical string theory,''
  Phys.\ Lett.\ B {\bf 428} (1998) 105
  [hep-th/9802109].
\bibitem{bpz}
 A.~A.~Belavin, A.~M.~Polyakov and A.~B.~Zamolodchikov,
  ``Infinite Conformal Symmetry in Two-Dimensional Quantum Field Theory,''
  Nucl.\ Phys.\ B {\bf 241} (1984) 333.
\bibitem{zam} 
A.~B.~Zamolodchikov,
  ``Higher Order Integrals of Motion in Two-Dimensional Models of the Field Theory with a Broken Conformal Symmetry,''
  JETP Lett.\  {\bf 46} (1987) 160
   [Pisma Zh.\ Eksp.\ Teor.\ Fiz.\  {\bf 46} (1987) 129].
 A.~B.~Zamolodchikov,
  ``Integrals of Motion in Scaling Three State Potts Model Field Theory,''
  Int.\ J.\ Mod.\ Phys.\ A {\bf 3} (1988) 743.
A.~B.~Zamolodchikov,
  ``Integrals of Motion and S Matrix of the (Scaled) T=T(c) Ising Model with Magnetic Field,''
  Int.\ J.\ Mod.\ Phys.\ A {\bf 4} (1989) 4235.
A.~B.~Zamolodchikov in 
``Integrable Systems in QFT and Statistical Mechanics,'' 
Adv. Stud. Pure Math. \textbf{19} (1989) 641
\bibitem{muss} 
G.Mussardo, 
``Statistical Field Theory: An Introduction to Exactly Solved Models in Statistical Physics'', Oxford University Press Inc., New York, 2010.
\bibitem{sz}
P.~Christe and G.~Mussardo,
  ``Integrable Systems Away from Criticality: The Toda Field Theory and S Matrix of the Tricritical Ising Model,''
  Nucl.\ Phys.\ B {\bf 330} (1990) 465.
P.~Christe and G.~Mussardo,
  ``Elastic s Matrices in (1+1)-Dimensions and Toda Field Theories,''
  Int.\ J.\ Mod.\ Phys.\ A {\bf 5} (1990) 4581.
G.~Sotkov and C.~-J.~Zhu,
  ``Bootstrap Fusions And Tricritical Potts Model Away From Criticality,''
  Phys.\ Lett.\ B {\bf 229} (1989) 391.
V.~A.~Fateev and A.~B.~Zamolodchikov,
  ``Conformal field theory and purely elastic S matrices,''
  Int.\ J.\ Mod.\ Phys.\ A {\bf 5} (1990) 1025.
D.~Bernard and A.~Leclair,
  ``Quantum group symmetries and nonlocal currents in 2-D QFT,''
  Commun.\ Math.\ Phys.\  {\bf 142} (1991) 99.
\bibitem{azz}
 A.~B.~Zamolodchikov,
  ``Exact Two Particle s Matrix of Quantum Sine-Gordon Solitons,''
  Pisma Zh.\ Eksp.\ Teor.\ Fiz.\  {\bf 25} (1977) 499
   [Commun.\ Math.\ Phys.\  {\bf 55} (1977) 183].
A.~B.~Zamolodchikov and A.~B.~Zamolodchikov,
  ``Factorized s Matrices in Two-Dimensions as the Exact Solutions of Certain Relativistic Quantum Field Models,''
  Annals Phys.\  {\bf 120} (1979) 253.
F.~A.~Smirnov, ``Form Factors in completely integrable models of Quantum Field Theory'', Cambridge University Press,1995.
\bibitem{bose}
A.~O.~Gogolin, A.~A.~Nersesyan and A.~M.~Tsvelik, ``Bosonization and Strongly Correlated Systems'', Cambridge University Press,1998.
\bibitem{1} 
 E.~A.~Bergshoeff, O.~Hohm and P.~K.~Townsend,
  ``Massive Gravity in Three Dimensions,''
  Phys.\ Rev.\ Lett.\  {\bf 102} (2009) 201301
  [arXiv:0901.1766 [hep-th]].
\bibitem{gub}
S.~S.~Gubser,
  ``Nonconformal examples of AdS / CFT,''
  Class.\ Quant.\ Grav.\  {\bf 17} (2000) 1081
  [hep-th/9910117].
\bibitem{VVB} 
J.~de Boer,
  ``The Holographic renormalization group,''
  Fortsch.\ Phys.\  {\bf 49} (2001) 339
  [hep-th/0101026].
J.~de Boer, E.~P.~Verlinde and H.~L.~Verlinde,
  ``On the holographic renormalization group,''
  JHEP {\bf 0008} (2000) 003
  [hep-th/9912012].
\bibitem{4}
 D.~Z.~Freedman, S.~S.~Gubser, K.~Pilch and N.~P.~Warner,
  ``Renormalization group flows from holography supersymmetry and a c theorem,''
  Adv.\ Theor.\ Math.\ Phys.\  {\bf 3} (1999) 363
  [hep-th/9904017].
 L.~Girardello, M.~Petrini, M.~Porrati and A.~Zaffaroni,
  ``Novel local CFT and exact results on perturbations of N=4 superYang Mills from AdS dynamics,''
  JHEP {\bf 9812} (1998) 022
  [hep-th/9810126].
 M.~Bianchi, D.~Z.~Freedman and K.~Skenderis,
  ``How to go with an RG flow,''
  JHEP {\bf 0108} (2001) 041
  [hep-th/0105276].  
\bibitem{2}
G.~Clement,
  ``Warped AdS(3) black holes in new massive gravity,''
  Class.\ Quant.\ Grav.\  {\bf 26}, 105015 (2009)
  [arXiv:0902.4634 [hep-th]].
 E.~Ayon-Beato, G.~Giribet and M.~Hassaine,
  ``Bending AdS Waves with New Massive Gravity,''
  JHEP {\bf 0905} (2009) 029
  [arXiv:0904.0668 [hep-th]].
J.~Oliva, D.~Tempo and R.~Troncoso,
  ``Three-dimensional black holes, gravitational solitons, kinks and wormholes for BHT massive gravity,''
  JHEP {\bf 0907} (2009) 011
  [arXiv:0905.1545 [hep-th]].
H.~Ahmedov and A.~N.~Aliev,
  ``The General Type N Solution of New Massive Gravity,''
  Phys.\ Lett.\ B {\bf 694} (2010) 143
  [arXiv:1008.0303 [hep-th]].
\bibitem{3}
H.~L.~C.~Louzada, U.~Camara dS and G.~M.~Sotkov,
  ``Massive 3D Gravity Big-Bounce,''
  Phys.\ Lett.\ B {\bf 686} (2010) 268
  [arXiv:1001.3622 [hep-th]].
\bibitem{nmg} 
U.~d.~.Camara and G.~M.~Sotkov,
  ``New Massive Gravity Domain Walls,''
  Phys.\ Lett.\ B {\bf 694} (2010) 94
  [arXiv:1008.2553 [hep-th]].
\bibitem{pos}
U.~Camara and G.~Sotkov,
  ``Geometry of the new massive gravity domain walls,''
  PoS ICFI {\bf 2010} (2010) 026.
\bibitem{8}
 O.~Hohm and E.~Tonni,
  ``A boundary stress tensor for higher-derivative gravity in AdS and Lifshitz backgrounds,''
  JHEP {\bf 1004} (2010) 093
  [arXiv:1001.3598 [hep-th]].
\bibitem{9}
J.~D.~Brown and M.~Henneaux,
  ``Central Charges in the Canonical Realization of Asymptotic Symmetries: An Example from Three-Dimensional Gravity,''
  Commun.\ Math.\ Phys.\  {\bf 104} (1986) 207.
\bibitem{cardy} 
J.~Cardy, ``Scaling and Renormalization in Statistical Physics,'' Cambridge Univ. Press,1996.
\bibitem{freed}
  D.~Bak, M.~Gutperle and S.~Hirano,
  ``A Dilatonic deformation of AdS(5) and its field theory dual,''
  JHEP {\bf 0305} (2003) 072
  [hep-th/0304129].
D.~Z.~Freedman, C.~Nunez, M.~Schnabl and K.~Skenderis,
  ``Fake supergravity and domain wall stability,''
  Phys.\ Rev.\ D {\bf 69} (2004) 104027
  [hep-th/0312055].
\bibitem{BF}
 P.~Breitenlohner and D.~Z.~Freedman,
  ``Positive Energy in anti-De Sitter Backgrounds and Gauged Extended Supergravity,''
  Phys.\ Lett.\ B {\bf 115} (1982) 197.
P.~Breitenlohner and D.~Z.~Freedman,
  ``Stability in Gauged Extended Supergravity,''
  Annals Phys.\  {\bf 144} (1982) 249.
\bibitem{x} 
  A.~B.~Zamolodchikov,
  ``Irreversibility of the Flux of the Renormalization Group in a 2D Field Theory,''
  JETP Lett.\  {\bf 43} (1986) 730
   [Pisma Zh.\ Eksp.\ Teor.\ Fiz.\  {\bf 43} (1986) 565].
A.~B.~Zamolodchikov,
  ``Renormalization Group and Perturbation Theory Near Fixed Points in Two-Dimensional Field Theory,''
  Sov.\ J.\ Nucl.\ Phys.\  {\bf 46} (1987) 1090
   [Yad.\ Fiz.\  {\bf 46} (1987) 1819].
\bibitem{sinha} 
 A.~Sinha,
  ``On the new massive gravity and AdS/CFT,''
  JHEP {\bf 1006} (2010) 061
  [arXiv:1003.0683 [hep-th]].
R.~C.~Myers and A.~Sinha,
  ``Seeing a c-theorem with holography,''
  Phys.\ Rev.\ D {\bf 82} (2010) 046006
  [arXiv:1006.1263 [hep-th]].  
\bibitem{fat}
 S.~L.~Lukyanov and V.~A.~Fateev,
  ``Physics reviews: Additional symmetries and exactly soluble models in two-dimensional conformal field theory,''
  Chur, Switzerland: Harwood (1990) 117 p. (Soviet Scientific Reviews A, Physics: 15.2)
 G.~Sotkov and M.~Stanishkov,
  ``Affine geometry and W(n) gravities,''
  Nucl.\ Phys.\ B {\bf 356}, 439 (1991).
\bibitem{zam-zam}
A.~B.~Zamolodchikov and A.~B.~Zamolodchikov,
  ``Structure constants and conformal bootstrap in Liouville field theory,''
  Nucl.\ Phys.\ B {\bf 477} (1996) 577
  [hep-th/9506136].

\end{thebibliography}
\end{document}